\documentclass[aps,12pt,preprint]{revtex4}

\providecommand{\LyX}{L\kern-.1667em\lower.25em\hbox{Y}\kern-.125emX\@}

\newcommand{\bee}{\begin{equation}}
\newcommand{\ee}{\end{equation}}
\newcommand{\beea}{\begin{eqnarray}}
\newcommand{\eea}{\end{eqnarray}}

\makeatother
\begin{document}
\newcommand{\ra}{\rightarrow}

\newcommand{\la}{\leftarrow}

\hspace*{10cm} COLO-HEP-498\\
 \hspace*{10.5cm} January 2004

\title{Moduli potentials in string compactifications with fluxes: mapping the Discretuum }

\author{R. Brustein\protect\( ^{(1)}\protect \),
S. P. de Alwis\protect\( ^{(2)}\protect \)}

\affiliation{(1) Department of Physics, Ben-Gurion University,
Beer-Sheva 84105, Israel
\\
(2) Department of Physics,  University of Colorado, UCB 390, Boulder, CO 80309, U.S.A.\\
 \texttt{e-mail:  ramyb@bgumail.bgu.ac.il },
\texttt{dealwis@pizero.colorado.edu}}

\begin{abstract}

We find de Sitter and flat space solutions with all moduli
stabilized in four dimensional supergravity theories derived from
the heterotic and type II string theories, and explain how all the
previously known obstacles to finding such solutions can be
removed. Further, we argue that if the compact manifold allows a
large enough space of discrete topological choices then it is
possible to tune the parameters of the four dimensional
supergravity such that a hierarchy is created and the solutions
lie in the outer region of moduli space in which the compact
volume is large in string units, the string coupling is weak, and
string perturbation theory is valid. We show that at least two
light chiral superfields are required for this scenario to work,
however, one field is sufficient to obtain a minimum with an
acceptably small and negative cosmological constant. We discuss
cosmological issues of the scenario and the possible role of
anthropic considerations in choosing the vacuum of the theory. We
conclude that the most likely stable vacuua are in or near the
central region of moduli space where string perturbation theory is
not strictly valid, and that anthropic considerations cannot help
much in choosing a vacuum.

\end{abstract}

\maketitle

\section{Introduction }

Time was when string theory was expected to produce a unique four
dimensional theory. However, even though in ten or eleven
dimensions superstring theory is perhaps unique in  that the
five perturbative formulations are expected to be different limits
of one underlying theory, it was realized already in the mid 80's
that there were many different possibilities for supersymmetric
vacua when the theory is compactified down to four dimensions.
 At that time it was thought that the only
phenomenologically viable theory was the heterotic theory, so it
was possible to argue that the gauge group had to be a subgroup of
\( E_{8}\times E_{8} \) or \( SO(32) \). Work done in the 90's
with F-theory removed even this constraint, so that now it appears
that one can find supersymmetric vacua with almost any gauge group
up to a rank of $O(1000)$, as well as many different numbers of
generations.

In order to get some perspective on the current state of the
theory it is useful to recall the steps which have led us to the
models in the so-called discretuum. We start with the basic
theoretical conjecture (T) and then add the different experimental
and observational inputs (E) that need to be used in order to get
a model of the real world.

\subsection{The saga of weakly coupled strings}

\begin{itemize}

\item T: Assumption of (weakly interacting) quantized superstrings
in a Lorentz invariant background. This yields a startling outcome
- the graviton (coupling precisely as expected from general
relativity at low energies) as well as the quanta of gauge fields,
thus giving us a viable candidate for a unified theory. However,
space time is ten-dimensional. Of course, since only two of these
dimensions need to be geometrical, the rest being contributions to
the central charge of a superconformal field theory, this leaves
open the possibility of a four dimensional space-time.

\item E1 - \emph{Four Dimensions}:

Since the observed world is four dimensional and this fact does
not emerge automatically from the basic conjecture, it has to be
put in as an extra assumption. Thus, the topological criterion
that the ten dimensional space is of the form \( R^{4}\times
M_{6}, \) $M_6$ being some compact manifold or some abstract
conformal field theory, was imposed, relegating all other
solutions of the theory including the simplest one \( R^{10} \) to
a theoretical limbo.

\item E2 \( \cal N\le  \) 1:

Simple, for instance toroidal, compactifications yield 16 (or 32)
supersymmetries in four dimensions which certainly cannot yield
the chiral structure of the observed world. Thus an additional
input, that only 4 supersymmetries survive, was added. So only
internal manifolds such as Calabi-Yau (CY) manifolds (including
their orbifold limits) were to be considered. A choice of such a
solution is characterized by a number of parameters - the \(
h_{12} \) complex structures and \( h_{11} \) Kahler structures of
the manifold. There are arguments that the space of such manifolds
is connected, though there may also be isolated points in this
so-called super moduli space that correspond to non-geometrical
compactifications such as asymmetric orbifolds.

\item E3 \emph{No massless moduli}:

The (super) moduli appear as massless four dimensional (chiral
super) fields in the low energy four dimensional action that is
supposed to describe the real world at scales below the string
scale. However they couple with gravitational strength to other
fields including the standard model ones, and such fields are
definitely ruled out by experiment since they affect Newtonian
gravity at large distances. The same is true of the dilaton
superfield whose ground state value sets the coupling strength.
Thus, additional input is needed which would generate a potential
for these moduli fields.

\item E4 \emph{Supersymmetry is broken:}

A moduli potential can be generated in many different ways. The
earliest solution was to consider gaugino condensation in a gauge
group\cite{Derendinger:1985kk,Dine:1985rz,Derendinger:1985cv} (See
\cite{Nilles:2004zg} for a recent review). Typically this yields a
runaway potential for the moduli but if there is a direct product
of gauge groups (obtained, say, by turning on discrete Wilson
lines in the internal manifold) then one has the possibility of
developing a critical point in the so-called ``race-track" models
\cite{Krasnikov:jj,deCarlos:1992da}. Similar effects can be
obtained by considering brane instanton effects
\cite{Witten:1996bn}. In addition, contributions to the potential
can be generated by turning on  fluxes in internal compact
directions \cite{Dine:1985rz}. Often the minima are supersymmetric
with a string scale negative cosmological constant (CC). However,
the world is not supersymmetric and at best has broken
supersymmetry (SUSY) with mass splitting at a scale \(
10^{-15}M_{P} \).

\item E5 \emph{The cosmological constant is small and positive}:

The world appears to have a positive (or perhaps zero) CC at a
scale \( 10^{-120}M_{P}^{4} \) . Recent work \cite{Bousso:2000xa}
has indicated that even though the natural scale of string theory
\( M\sim M_{P} \) it may still be possible to find a small
positive CC.

\item E6, \& E7 \emph{Three generations and \(
SU(3)\times SU(2)\times U(1) \)} : It may turn out that a string
theory which satisfies all of the above criteria will be unique
and result in three generations of chiral fermions with just the
standard model gauge group. However, given the enormous number of
possibilities this seems unlikely and so we may need these two
experimental inputs as well.
\end{itemize}

\subsection{The Discretuum}

What then is left for string theory to predict? It is likely that
the Yukawa couplings can be calculated once a model satisfying all
the other criteria is found. Here, again, although the
superpotential terms can be calculated exactly and will not be
renormalized, the Kahler terms will acquire corrections, but
perhaps these can be calculated perturbatively to a reasonable
accuracy. However, even if this succeeds it is clear that we are
far from a fundamental theory involving just some basic
theoretical input(s).

Now the question arises as to what to make of the huge number of
solutions that do not satisfy the observational inputs E1-E7. Do
they exist as different universes? For instance is there a ten
(non-compact) dimensional supersymmetric universe? If one only
uses E1 then there are a continuous infinity to the power of the
dimension of moduli space, of supersymmetric compactifications
corresponding to values of the moduli. These are perfectly valid
solutions of perturbative string theory. Are they all to be
included as Universes that actually exist? E2 selects a subset of
the above but E4 and E5 give a new set that comes from giving a
potential to the moduli \footnote{Except when the minimum of the
potential - the CC is exactly zero in which case what is selected
is a discrete subset of the set of vacua obtained after E2.}. It
is this discrete set that is referred to in the literature as the
`Discretuum'.

In supergravity (SUGRA) phenomenology after picking the field
content and the gauge group one is left with three arbitrary
functions. A realistic phenomenomnology with a zero or a tiny CC
could be obtained by fine tuning. In string theory there was no
obvious mechanism that would allow for this fine tuning, let alone
finding a solution to the problem of generating a constant which
is 120 orders of magnitude smaller than the string scale. Our
interpretation of the work of  Bousso and Polchinski
\cite{Bousso:2000xa} is that such fine tuning is actually possible
in string theory in spite of the quantization of the parameters
(fluxes) in terms of the string scale.

In this paper we will consider both heterotic and type IIB weakly
coupled string theories compactified on large volume 6D manifolds,
and examine under what conditions one can get a small positive CC
with hierarchically larger, but still parametrically small in
string units, SUSY breaking. We will attempt to find such minima
from the F-terms of the \( \cal N \)=1 potential in race-track
type models. We feel that finding solutions for which SUSY is
spontaneously broken through F-terms is more reliable than
invoking either explicit breaking \cite{Kachru:2003aw} (KKLT) or
D-terms \cite{Burgess:2003ic}. If there is only one light modulus
as in the KKLT case we find that it is not possible to find a SUSY
breaking minimum with zero or positive CC. However, with at least
two light moduli such minima do exist. In general, we find that
each of the vacua in the discretuum develops its own discretuum
with several vacua, including some that have negative CC and
unbroken SUSY. With one light modulus, however, it is possible to
find a minimum in which SUSY is broken and the CC is negative
though acceptably small.

We have not yet produced a concrete string model which realizes
all the constraints. However, we do show that all known obstacles
that were previously found in race-track models (see, for example,
\cite{Dine:1999dx}) can be removed in this framework by a choice
of topological or geometrical properties of the compactification
manifold. We explain along the way what  the obstacles are and why
previous attempts failed.

We will argue that all solutions in the discretuum that are in the
outer region of moduli space, including ours, are not
cosmologically viable - being subject to the overshoot problem
first discussed in \cite{Brustein:1993nk} and recently called
appropriately the ``bat from hell" problem \cite{dinepc}. We also
discuss the possible application of the anthropic principle to
choose among the variety of vacuua and find that it is not very
useful. We expect that solutions in the central region of moduli
space will not suffer from the cosmological overshoot problem.

\section{Moduli potential in the heterotic string theory}

\subsection{The potential of the complex structure moduli and the
dilaton}

The first attempt at using fluxes and gaugino condensation to
stabilize the moduli was that of Dine et al \cite{Dine:1985rz}
(for a recent discussion of this model see \cite{Gukov:2003cy}).
The main argument for rejecting this model as a model of moduli
stabilization had been the observation that flux is quantized in
string units \cite{Rohm:1986jv}, and hence the dilaton is
stabilized in an unrealistic strong coupling region.
 Here we revisit the argument and show that
it should be modified, and that it is possible to stabilize
the dilaton at weak (or intermediate) coupling. This, as far as we know, could
have been observed at the time that the original paper was written.

The ten dimensional low-energy effective action is reduced to a
four dimensional action using the ansatz,\[
ds_{10}^{2}=e^{-6u(x)}ds_{4}^{2}+e^{2u(x)}g_{mn}^{0}dy^{m}dy^{n},\]
where \( m,n \) go over the dimensions of the internal space which
is taken to be a CY 3-fold \( X \). For simplicity we will assume
that \( X \) has only one Kahler modulus but may have an arbitrary
number of complex structure moduli. The four dimensional dilaton
\( \varphi  \) and the four dimensional volume scalar \( \rho  \)
are related to the ten dimensional dilaton \( \phi  \) and the
modulus \( u \) by \( \varphi =\frac{\phi }{2}-6u \) and \( \rho
=\frac{\phi }{2}+2u \) . The chiral superfields \( S,T \) are then
defined by \( S=e^{-\varphi }+ia \) and \( T=e^{\rho }+ib \) where
\( a,\, b \) are the corresponding axions. The argument proceeds
from the observation that the low energy ten dimensional effective
action for the heterotic string contains the following
contribution that can be interpreted as an effective potential in
four dimensions for the moduli \footnote{In this section we use
the same normalization conventions for the classical action as in
\cite{Gukov:2003cy}.}
 \begin{equation}
 \label{vaction}
V_{action} = \frac{1}{4\alpha'^{4}}\frac{1}{S_{R}T_{R}^{3}}
\int_{X} \left(H_{3} - \frac{\alpha
'}{16}T_{R}^{3/2}S_{R}^{3/2}T_{3}\right)\wedge
^{*_{6}}\left(H_{3}-\frac{\alpha
'}{16}T_{R}^{3/2}S_{R}^{3/2}T_{3}\right).
 \end{equation}
In the above \( H_{3} \) is the NSNS three form flux which is
taken to be non-zero only on \( X \) and \( T_{3} \) is a
fermionic bilinear three form (not to be confused with the chiral
superfield $T$) which is assumed to be represented upon gaugino
condensation by
\begin{equation}
\label{T3} T_{3}=2U\Omega +c.c,
\end{equation}
where \( U=\langle tr\lambda \lambda \rangle \) is an effective
low energy scalar field representing the gaugino condensate and \(
\Omega  \) is the holomorphic (3,0)-form on \( X \)
\cite{Dine:1985rz}. $S_R$, $T_R$ stand for the real parts of $S$,
$T$, respectively. The dynamics of \( U \) is governed by the
Veneziano-Yankielowicz (VY) superpotential \cite{Veneziano:1982ah}
\begin{equation}
 \label{vy}
W_{np} =\frac{U}{4}\left[f+\frac{C(G)}{8\pi^{2}}
 \ln \left(\alpha '^{\frac{3}{2}}U\right)\right],
\end{equation}
 where \( f \) is the gauge coupling function and \( C(G) \) is
the dual Coxeter number of the gauge group which we have assumed
here to be simple. We have also assumed that the model does not
contain matter that is charged under the gauge group. Classically
\( f=S \), so extremizing the VY effective superpotential one
finds \( U=\alpha '^{-3/2}e^{-\frac{8\pi ^{2}}{C(G)}S-1} \)
\footnote{Note that the precise normalization depends on the
cutoff scale chosen in (\ref{vy}). Here we have chosen it to be the
string scale.}. We also note for future reference that the
non-perturbative superpotential for the modulus \( S \) that is
generated is
\begin{equation}
 \label{wnp}
 W_{np}=-C(G)\mu ^{3}e^{-\frac{8\pi^{2}}{C(G)}S-1},
\end{equation}
where \( \mu ^{3}=\frac{1}{32\pi ^{2}\alpha '^{3/2}} \) .

Apart from  Chern-Simons terms which are \( O(\alpha ') \)
corrections, \( H_3 \) is closed and the {\it classical}  equation
of motion implies that it is co-closed as well. Thus it  may be
expanded in terms of a basis of harmonic three forms on \( X \),
\begin{equation}
 \label{expansion}
H= a\Omega +b^{\alpha }\chi _{\alpha}+\overline{a}\overline{\Omega
} + \overline{b}^{\overline{\beta }}\overline{\chi
}_{\overline{\beta }},
\end{equation}
where the sums over \( \alpha ,\overline{\beta } \) go over
\(1,...,h_{12} \), the dimension of the complex structure moduli
space of \( X. \) The volume of the CY manifold can be expressed
in terms of $\Omega$ by $v\equiv i\int \Omega \wedge
\overline{\Omega }$, and the metric on the moduli space is given
by $G_{\alpha \overline{\beta }}=- \frac{i}{v} \int \chi _{\alpha}
\wedge \chi _{\overline{\beta }} $. Using expansion
(\ref{expansion}) and the expression for \( T_{3} \) in
eq.(\ref{vaction}), we get
\begin{equation}
 \label{vaction2}
 V_{action}=\frac{1}{2\alpha'^{3}} \frac{v} {S_{R}T_{R}^{3}}
 \left[\left|a- \frac{\alpha '}{8} \frac{3\pi^{2}}{C_{G}}
 T_{R}^{3}S_{R}^{1/2}W_{np}\right|^{2} + G_{\alpha \overline{\beta
}}b^{\alpha }b^{\overline{\beta }}\right].
\end{equation}

We now wish to express the action (\ref{vaction2}) in the \( N=1
\) SUGRA form
\begin{equation}
 \label{vsugra}
V_{SUGRA}=e^{K}\left(K^{i\overline{j}}F_{i}F_{\overline{j}}-3|W|^{2}\right).
\end{equation}
From the classical action and the properties of the manifold \( X
\) the Kahler potential is found to be
\begin{equation}
\label{kahler1} K=-\ln (S+\overline{S})-3\ln (T+\overline{T}) -
\ln\left(\frac{v}{4\alpha '^{3}}\right).
\end{equation}
The superpotential is the sum of two contributions. One
contribution comes from the flux superpotential of Gukov, Vafa and
Witten \cite{Gukov:1999ya} which is given by
\begin{equation}
 \label{wflux}
 W_{flux}=\frac{4}{\alpha '^{4}}\int H\wedge \Omega = -
\frac{4 i} {\alpha '^{4}} a v.
\end{equation}
To obtain the second equality we have used the expansion
(\ref{expansion}). In the case that $W_{np}$ vanishes, $W_{flux}$
and the Kahler potential (\ref{kahler1}) result in the potential
coming from the \emph{classical} action (\ref{vaction}) with \(
W_{np} \) set to zero. Now, if gaugino condensation does occur and
$W_{np}$ does not vanish, the total superpotential is the sum of
the flux superpotential (\ref{wflux}), and the gaugino condensate
superpotential given by (\ref{wnp}),
\[ W_{tot}=W_{flux}+W_{np}.\]
Computing (\ref{vsugra}) with $W_{tot}$ gives
\begin{eqnarray}
V_{SUGRA}=  && \frac{\alpha '^{4}} {32T_{R}^{3}S_{R} v} \Biggr\{
\left| \frac{4i} {\alpha '^{4}} \overline{a} v
+ \left(1 + \frac{16\pi ^{2}}{C(G)} S_{R}\right) W_{np}\right|^{2}  \nonumber \\
+ && G^{\alpha \overline{\beta }}\left(\frac{4i}{\alpha '}
b^{\overline{\gamma}} G_{\alpha \overline{\gamma
}}v+\partial_{\alpha} K W_{np} \right) \left(\frac{4i}{\alpha '}
b^{\overline{\beta }} G_{\delta \overline{\beta }} v +
\partial_{\overline{\beta }}K W_{np}\right) \Biggl\}.
 \label{vsugra2}
\end{eqnarray}
Here the differentiation $\partial_{\alpha}$ is with respect to the complex
structure moduli.

Comparing (\ref{vaction2}) to (\ref{vsugra2}) we see that there is
agreement only when we set \( W_{np}=0 \), i.e., only at the
classical level \footnote{Aspects of this difference have been
noticed already in \cite{Dine:1985rz}.}. This should not be
surprising. One should not expect to obtain the correct
non-perturbative four dimensional action from the \emph{classical}
ten dimensional action. The difference between (\ref{vaction}) and
(\ref{vsugra2}) is significant. If (\ref{vaction}) had been the
correct formula for the potential then it would be impossible to
find an  \( O(1) \) solution for the four dimensional dilaton \( S
\) and hence a weak 4D gauge coupling . This follows from
integrating the relation between \( H \) and \( T_{3} \) that is
obtained at the minimum of \( V_{action} \) in (\ref{vaction})
over a three cycle on \( X \) and using (\ref{T3}) for \( T_{3} \)
and the quantization of the three form field \( H=dB \) that was
first observed in \cite{Rohm:1986jv} (for a recent discussion see
\cite{Gukov:2003cy}). However, this is not the correct relation at
the minimum since the \( \cal N \)=1 SUGRA potential is
(\ref{vsugra2}) and the correct equation is the vanishing of the
\( F \) term for the dilaton,
 \begin{equation}
 \label{Fs}
\frac{4i}{\alpha '^{4}}\overline{a}v
+(1+\frac{16\pi^{2}}{C(G)}S_{R})W_{np}=0.
\end{equation}

Substituting the explicit expression for $W_{np}$ from
eq.(\ref{wnp}) we get,
\begin{equation}
 \label{weakcoupling}
\frac{4i}{\alpha '^{4}}\overline{a}v -  \frac{C(G)}{32 \pi ^{2}
\alpha'^{3/2}} \left(1+\frac{16\pi^{2}}{C(G)}S_{R}\right)
e^{-\left[\left(\frac{8\pi^{2}}{C(G)}\right)S+ 1\right]}=0.
\end{equation}
Thus, getting a weak coupling solution with $S_R$ of order a few
amounts to finding a small value in the ``discretuum'' for the
flux superpotential, specifically for the product $av$. This is
similar to the corresponding type IIB case discussed by KKLT, and
as in that case, one expects such values to exist in CY manifolds
with large numbers of complex structures. This mechanism would be
an alternative to the proposal of \cite{Gukov:2003cy} where the
Chern-Simons contributions to $H$ \cite{Derendinger:1985cv} were
included and integrated over spaces where the corresponding
invariants are fractional.\footnote{In this reference it was
argued that one is forced to do this - based on the constraint
mentioned above from using the form of the potential coming from
the ten dimensional action. Here we have seen that this constraint
is not the appropriate one.}

In addition to (\ref{Fs}), at the potential minimum the $F$ term
of the complex structure moduli needs to vanish,
 \begin{equation}
 \label{compstruc}
\frac{ 4 i } {\alpha'} b^{\overline{\gamma }}G_{\alpha
\overline{\gamma }}v + \partial_{\alpha}K W_{np}=0.
\end{equation}
Using (\ref{Fs}), and the relation $b^{\overline{\gamma
}}G_{\alpha \overline{\gamma }}=b_{\alpha }$, we get,
\begin{equation}
\label{compstruc2}
\partial_{\alpha}K=\alpha'^3 \left(1+(16\pi ^{2}/C(G))S_{R}\right)
\frac{b_{\alpha }}{\overline{a}}
\end{equation}
Recall that the derivation here is with respect to the complex
structure moduli. Thus, with generic fluxes the potential
(\ref{vaction2}) fixes all the complex structure moduli in
addition to the dilaton \( S \). Note that unlike in the case of
type IIB, here in order to get a solution we need the
non-perturbative term \( W_{np} \). From (\ref{compstruc2})
 it is clear that the relation
  \(b_{\alpha }=0 \) imposed in \cite{Gukov:2003cy} (and would have been
obtained if we had used (\ref{vaction2}) rather than
(\ref{vsugra2})) is valid only if \( a=0 \) or \(
S_{R}\rightarrow \infty  \).

\subsection{The Potential of the Kahler moduli}

The Kahler moduli, and in particular the volume modulus \( T \)
which is present in any compactification, are clearly not fixed in
the models that we have discussed so far. Additionally, SUSY is
generically  broken since
\begin{equation}
\label{Ft} F_{T}=K_{T}W= - \frac{3}{2T_{R}} \left(\frac{4i}{\alpha
'^{4}}\overline{a}v+W_{np}\right)= \frac{6i}{T_{R}\alpha '^{4}}
\frac{2+\frac{16\pi ^{2}}{C(G)}S_{R}}{1+\frac{16\pi
^{2}}{C(G)}S_{R}}\overline{a}v
\end{equation}
is non-zero for a finite value of $T_R$ and generic fluxes. In
fact unbroken SUSY (\( F_{T}=0 \)) occurs only in
the decompactification limit $ T_{R}\rightarrow \infty$,
 as long as the flux superpotential (in effect \( a \)) is non-zero.
The situation here is somewhat different from that in type IIB,
where \( S \) and the complex structure moduli were fixed
classically i.e. without any non-perturbative superpotentials
\cite{Giddings:2001yu} and where even though generically SUSY was
broken there were flux configurations which preserved it.

We would like to consider possible modifications of the models so
that they will stabilize all moduli, including the Kahler moduli.
For simplicity we discuss now the case of only one Kahler modulus,
the volume.

A dependence on the \( T \) modulus can arise from threshold
effects which have been calculated for various compactifications.
We will first consider the \( T \) dependent contribution to the
gauge function coming from anomaly considerations . Then we will
make some remarks about the case when the compact manifold \( X \)
is an orbifold where the complete one-loop string theory
correction has been worked out.

There are now two possibilities for analyzing the theory with all
moduli stabilized.

\begin{itemize}

 \item
The first possibility is to integrate out the complex structure
moduli and the dilaton by arguing that they are fixed at the
minimum of the potential (\ref{vsugra2}) at a high scale.

This would generically be the case. In fact without threshold
corrections (i.e. just using the classical relation \( f=S \)) the
\( T \) modulus would have zero mass, and the other moduli would
have string scale masses \footnote{A caveat noticed by Michael
Dine is that the mass matrix of moduli is a large matrix, and
therefore it can have some particularly small or large
eigenvalues.}. In this respect the situation is similar to the
type IIB case. Then by incorporating threshold corrections, we get
a theory for the single light field \( T \) with a constant
superpotential: (\( W_{tot} \) evaluated at the minimum of the
potential . As we have shown elsewhere \cite{Brustein:2001ci} the
modulus in the SUSY breaking direction needs to be light but there
is no such requirement on the other moduli.

 \item
The second possibility is to integrate out the complex structure
moduli at a high scale but arrange by a choice of a point in the
discretuum to have the dilaton light so it is not integrated out
at this stage.

It is plausible that within the discretuum there are choices of CY
manifolds and fluxes where this is justified. So we solve the
equations \(
\partial _{\alpha }V=0 \) to express the complex structure moduli
in terms of \( S \) and \( T \). It is not obvious that this can
be done holomorphically but since SUSY cannot be broken by this
procedure it must be the case that the result is a \( \cal N \)=1
SUGRA with just \( S \) and \( T \) moduli. Then after including
threshold corrections (that would introduce \( T \) dependence in
the superpotential) we would be left with a two-moduli
minimization problem. Since now we don't have a positive definite
potential, finding actual solutions is complicated but can be
done. However, in this case one would in general expect the Kahler
potential to be different from the naive classical form obtained
by just suppressing the complex structure dependent part of $K$.

 \item
The complete minimization problem (if none of the moduli are
integrated out at a high scale) is prohibitively complicated. But
as an alternative to the previous possibility we can consider the
case where \( h_{12}=0 \) as in the original paper
\cite{Dine:1985rz}. In this case obviously we cannot use
cancellations among different 3 cycles to get a small value for \(
W_{flux} \) and we would have to resort to the mechanism of
\cite{Gukov:2003cy}, which uses the Chern-Simons terms with the
classical Kahler potential  to get realistic examples with two
light moduli.
\end{itemize}

We will show in section IV, that within models with only one light
modulus it is impossible to get a true minimum of the potential
with a zero or positive CC from the F-terms. With two light moduli
we have found examples where such minima exist. Thus, only the
last two cases will lead to models with all moduli stabilized with
a positive or zero CC.

\subsubsection{Threshold corrections from Green-Schwarz terms}

Regardless of the compactification manifold there are some
one-loop corrections that can be computed due to the existence of
the Green-Schwarz anomaly cancellation mechanism. As pointed out
by Banks and Dine \cite{Banks:1996ss} (see also
\cite{Gukov:2003cy} for a recent discussion) from reduction of the
\( B\wedge X_{8} \) term in the ten dimensional action (where \(
X_{8} \) is a certain polynomial in the gauge and curvature two
forms) it is possible to see that the gauge coupling function(s)
takes the form \( f_{i}=S+\beta _{i}T \) where the \( \beta _{i}
\) are numbers determined by the topology of the gauge bundle and
the tangent bundle of \( X \) . Let us also assume that \( X \) has a
non-trivial fundamental group so that we can turn on discrete
Wilson lines to break the original ten dimensional group \(
E_{8}\times E_{8} \) or \( SO(32) \), to a product of several
simple groups. This adds another layer of discrete choices in the
discretuum. Then the superpotential arising from gaugino
condensation (\ref{wnp}) takes the form
 \begin{equation}
 \label{wnp2}
 W_{np}=-\sum _{i}C(G_{i})\mu ^{3}e^{-\frac{8\pi ^{2}}{C(G_{i})}(S+\beta _{i}T)-1}.
\end{equation}

The potential is then given by (\ref{vsugra2}) with the above
expression for \( W_{np} \) and a term
\begin{equation}
 \label{dvsugra}
\Delta V_{SUGRA}=\frac{\alpha'^{4}}{32T_{R}^{3}S_{R}
v}\frac{4T_{R}^{2}}{3}\left\{(|\partial
_{T}W_{np}|^{2}-\frac{3}{T_{R}} Re(\partial
_{T}W_{np}(\overline{W}_{flux}+\overline{W}_{np})\right\}.
\end{equation}
If the fluxes, gauge group parameters and the various topological
numbers, are such that the dilaton and the complex structure are
heavy then they can be integrated out by setting the expression
(\ref{vsugra2}) to zero, and then effectively we have a potential
for the modulus \( T \) given by (\ref{dvsugra}) with \( S \) and
the complex structure moduli fixed. As we have already mentioned,
in this case with only one remaining light modulus it is not
possible to find de Sitter (dS) or Poincare minima. The general
situation is of course prohibitively complicated. Thus we will
focus on a situation where only the complex structure moduli
(assumed heavy) are integrated out leaving two light moduli,
$S,~T$. Alternatively,  we could have considered a manifold $X$
with two Kahler moduli with $S$ and the complex structure moduli
integrated out at a high scale.

\subsubsection{Modular invariance for orbifolds}

Another source of \( T \) dependence (at least in orbifold
compactifications) comes from requiring modular invariance under
\( M:T\rightarrow \frac{aT-ib}{icT+d},\, \, a,..d\in \, \cal Z \)
\cite{Font:1990nt} as might be expected from T-duality (of course,
in this case there would be three Kahler moduli but for the sake
of simplicity we will identify them).

Then (assuming that the complex structure moduli have been
integrated out at a high scale) we take
\begin{eqnarray}
\label{khet}
K && =-3\ln (T+\overline{T})-\ln (S+\overline{S})  \\
  W && =\left(c+\sum d_{i}e^{-8\pi
^{2}S/C(G_{i})}\right)/\eta (T)^{6}\equiv \omega (S)/\eta (T)^{6}
 \label{whet}
\end{eqnarray}
where \( \eta (T)=e^{-\pi T/12}\prod\limits_{n}\left(1-e^{-2\pi
nT}\right) \) is the Dedekind eta function. We have assumed here
that the Kahler potential is  equal to its classical form. The constant
in the superpotential arises from H-flux as in (\ref{wflux}). The
Kahler invariant combination of \( K,W \) and hence the potential
is \( M \) invariant.

The  potential resulting from $K$ and $W$ of eqs.(\ref{khet}),
(\ref{whet}) is
\begin{equation}
 \label{TdualV}
 V=\frac{\left|\eta (T)\right|^{-12}} {2S_{R}(2T_{R})^{3}} \left\{\left|2S_{R}\omega _{S}-\omega
\right|^{2}+ \left( \frac{3T_{R}^{2}} {\pi^{2}}
\left|\widehat{G}_{2}\right|^{2}-3\right)|\omega|^{2}\right\},
\end{equation}
where \( \widehat{G}_{2}=-\left(\frac{\pi }{T_{R}}+4\pi \eta
^{-1}\frac{\partial \eta }{\partial T}\right) \) is a modular
function of weight two. The potential has SUSY extrema
at\begin{eqnarray}
2S_{R}W_{S}-W(S) & =0 & \label{susy1} \\
\widehat{G}_{2} & =0. & \label{susy2}
\end{eqnarray}
With an appropriate choice of value of \( c \) in the
superpotential (\ref{whet}) there would be a solution for \( S \)
perhaps even at weak coupling and the corresponding Hessian in the
$S$ direction is positive definite as discussed in section 4.
However, the zeros of \( G_{2} \) (i.e. $ T=1,\, e^{i\pi /6}, 0$)
are saddle points (\( T=1 \)) or maxima in the $T$ direction. In
addition, there is a true minimum, (again, a result of a numerical
calculation) at T=1.2 independently of the value of S at the
minimum. At this point the volume of the compact manifold is not
large in string units, and hence we may expect large $\alpha'$
corrections to this, and the solution is not under complete
control.

\subsubsection{Threshold corrections for orbifolds}

It is not clear that in the presence of fluxes the theory is
modular invariant. In fact, if one strictly follows the logic as
in the type IIB case (discussed by KKLT)
 what one gets is a superpotential
\[
W=c+\sum d_{a}e^{-3kS/2\beta _{a}})/\eta (T)^{6}\equiv c+\omega
(S)/\eta (T)^{6},
\]
where \( c=\int H\wedge \Omega  \) evaluated at the minimum of the
classical flux potential at which generically all complex
structure moduli will be fixed. In computing the gaugino condensate from
(\ref{vy}) we have used the (Wilsonian - hence holomorphic) gauge
coupling function \[ f=kS+\frac{1}{4\pi
^{2}}\left(\frac{1}{2}b'-k\delta ^{GS}\right)\ln \eta (T)^{2}.\]
which comes from the calculation of threshold effects in orbifolds
\cite{Kaplunovsky:1995jw}. In contrast to this, the \( T \)
dependence of the first term in \( W \) in eq.(\ref{whet}) comes
from the \emph{requirement} of \( M \) invariance in the
potential.

However, now (with the same \( K
\) as before) there is no modular invariance. The
potential is
\[
V=\frac{|\eta (T)|^{-12}}{2S_{R}(2T_{R})^{3}}\left\{
\left|2S_{R}\omega _{S}-\omega -c\eta ^{6}(T)\right|^{2} + 3
\left|\frac{T_{R}^{}}{\pi ^{}}\widehat{G}_{2}\omega +c\eta
^{6}(T)\right|^{2}-3 \left|\omega  + c
\eta^{6}(T)\right|^{2}\right\}.
\]
This potential for two moduli is we believe the correct replacement
of the formula (\ref{TdualV}). It is a potential for two (possibly
light) moduli and we will discuss the minima of such potentials in
section 4.

\subsubsection{Non-Kahler manifolds}

It is well known  that in the presence of \( H \) flux the
heterotic string does not admit supersymmetric compactifications
on Kahler manifolds \cite{Strominger:1986uh}, \footnote{This can
be seen from the discussion after eq.(\ref{Ft}).}. Such
compactifications are possible, however, on non-Kahler manifolds
and recently there have been a number of papers on this subject
(see for example \cite{Cardoso:2003sp}, \cite{Becker:2003sh} and
references therein). The Non-Kahler manifolds are not Ricci flat
and so there are in general two contributions to the classical
potential - one from the fluxes and one from the curvature.
Actually, one might think that this is the case even if the
internal space is taken to be conformally CY. But in that case, as
we will see below, there is no solution unless the conformal
factor is trivial and the flux is zero.

The metric of the ten dimensional space is parametrized as
\begin{equation}
\label{metric3} ds^{2}= e^{2\omega (y)-6u(x)}
\tilde{g}_{\mu\nu}(x)dx^{\mu }dx^{\nu } +
 e^{-2\omega (y)+2u(x)}\tilde{g}_{mn}(x,y)dy^{m}dy^{n}.
\end{equation}
Then from the ten dimensional heterotic action we have the \emph{classical}
potential
 \[
 V = -\frac{1}{S_{R}}\int d^{6}y \sqrt{\widetilde{g}^{(6)}(y)}
 \left[\frac{1}{T_{R}^2}\left(\widetilde{R}^{(6)}-
 8\left(\widetilde{\partial} _{m}\omega \right)^{2}\right) -
 \frac{e^{4\omega (y)}}{12 T_{R}^{3}}\widetilde{H}_{3}^{2}\right].
  \]
This potential is a runaway potential in \( S \) and some quantum
(or stringy) effect such as the gaugino condensate term discussed
earlier is needed to stabilize it. On the other hand for \( T \)
the situation is different. If  \( \int d^{6}y
\sqrt{\widetilde{g}^{(6)}(y)} \widetilde{R}^{(6)}> 8 \int d^{6}y
\sqrt{\widetilde{g}^{(6)}(y)} \left(\tilde{\partial}_{m} \omega
\right)^{2} \), then it seems that it is possible to classically
stabilize the \( T \) modulus. Of course, if \( \tilde{g} \)
metric is CY as assumed in the previous subsections, then
$\widetilde R=0$ and there is no extremum point for \( T \)
either.

It has been suggested that the potential (in the non-Kahler case)
can be expressed in terms of a superpotential \( W=\int
(H-idJ)\wedge \Omega  \) where $J$ is the Kahler form (\cite{Cardoso:2003sp},
\cite{Becker:2003sh} and references therein). However it is not
clear what the Kahler potential is, and it is not known how to
express the potential coming from the ten-dimensional action in
the \( \cal N \) =1 SUGRA form. Once this is done then we will
have a situation which is dual to the type IIB case of GKP with
the \( S \) modulus being exchanged with the \( T \) modulus and
the former being stabilized by invoking non-perturbative effects
\footnote{We mention in passing that in the above mentioned
references it is argued that the $ T $ modulus stabilization is a
stringy effect needing the incorporation of $ \alpha ' $
corrections. We are somewhat puzzled by this since the above
argument does not require any such corrections.}. We leave further
discussion of these issues to future work.

\section{Moduli potential in type IIB string theory}

In type IIB string theory it was shown \cite{Giddings:2001yu}
(GKP) that all the complex structure moduli and the dilaton can be
stabilized by an appropriate choice of fluxes. The resulting 4D
models are of the no scale type. An important question is whether
the Kahler moduli can be stabilized with SUSY broken, in a
Poincare or dS background.

\subsection{Review of the proposals to stabilize Kahler moduli}

Let us first briefly review the proposals of KKLT and
\cite{Burgess:2003ic} (BKQ). Both of these constructions start
with the potential for the complex structure moduli and the
dilaton given in GKP \footnote{We use the same conventions as GKP
in this section.}. This is a classical N=1 SUGRA potential which
can be obtained by considering ten-dimensional low energy type IIB
theory compactified on a CY orientifold with \( D3 \) branes and
\( D7 \) branes - essentially a limit of an F-theory construction.

The metric is taken to be (\ref{metric3}) where \( e^{4u}=T_R \)
is the real part of the Kahler modulus (volume modulus) which sets
the overall size of the internal space and \( e^{\omega } \) is a
warp factor which effectively changes the scale of four
dimensional physics at different points on the internal manifold.
Additionally, we impose the constraint \( \partial _{\mu }\det
\tilde{g}_{mn}=0 \) on the \( \tilde{g}_{mn} \) metric which can
in turn be parametrized in terms of the other Kahler moduli as
well as the complex structure moduli. GKP considered the case
where ten dimensional space is compactified on a CY manifold with
only one Kahler modulus but an arbitrary number of complex
structure moduli. They derived a potential for these moduli and
the dilaton \footnote{Strictly speaking this derivation is valid
only when the warp factor is trivial \cite{deAlwis:2003sn} - the
solutions though are valid even for non-trivial $ \omega  $. }
\begin{equation}
 \label{potential}
 V=\int_{X}d^{6}y\sqrt{\tilde{g}^{(6)}}\frac{e^{4\omega
(y)-12u(x)}}{24 \tau _{I}}
\widetilde{\left|iG_{3}-*_{6}G_{3}\right|}^{2}.
\end{equation}
Here \( G_{3}=F_{3}-\tau H_{3} \), \( H_{3} \) is the NS three
form flux of type IIB, $F_{3}$ is the RR three form flux, and \(
\tau =C_{0}+ie^{-\phi } \) is the complex axion dilaton field.
The integration is over the CY manifold $X$, and we
have set \( 2\kappa ^{2}_{10}=1 \). The tilde over the absolute
value means that the Hodge dual and
the tensor contractions are evaluated with
the metric $\widetilde{g}_{mn}$. The potential (\ref{potential})
can be derived using the standard SUGRA form from a Kahler potential \( K \)
and a superpotential \( W \)   given by

\begin{eqnarray}
K= && -\ln [-i(\tau -\overline{\tau })] - 3\ln (T+\overline{T}) -
\ln\left(i\int_{X} \Omega \wedge \overline{\Omega }\right)
 \label{kahler} \\
W= && 8\int\limits_{X}G_{3}\wedge \Omega,
 \label{w0}
\end{eqnarray}
where \( \Omega  \) is the holomorphic three form on the CY
manifold \( X \) . The potential is clearly positive definite and
is of the no-scale form: at the minimum of the complex structure
moduli and the dilaton the potential vanishes so that the scale \(
T \) is undermined. At this point the fluxes must satisfy the
imaginary self-duality condition \( iG_{3}=*_{6}G_{3} \) and SUSY
is broken if \( W=c\ne 0 \) at this point.

To stabilize the volume modulus one may introduce non-perturbative
contributions to the superpotential, coming for instance from
gaugino condensation in the gauge theory on the stack of \( D7 \)
branes wrapping a four cycle (with betti number \( b_{1}=0 \)) in
the internal manifold, as suggested by KKLT. In this case it is
easily seen that the corresponding gauge coupling function of the
super-Yang-Mills theory is given by \( f=T \) so that by standard
arguments (reviewed in the previous section) a superpotential for
this modulus is generated - thus giving a total superpotential
\begin{equation}
 \label{wtot}
 W=c-C(G)\mu^{3}e^{-8\pi ^{2}T/C(G)}.
\end{equation}
With the (classical) Kahler potential (\ref{kahler}) the resulting
potential has a single negative minimum at which SUSY is
preserved, rather than being broken as before.

The KKLT proposal is to add the contribution of a \(
\overline{D}_{3} \) brane to this four dimensional effective
action. The anti-D brane gives a positive contribution to the
potential,
\[ \delta V=\frac{D}{T_{R}^{3}},\]
where \( D \) is positive and proportional to the \(
\overline{D}_{3} \) tension. KKLT add the \( \overline{D} \) brane
to the 4 dimensional effective action, however the \( \overline{D}
\) branes, like the \( D \) branes are string theoretic objects
and should be added to the classical ten dimensional theory. There
does not seem to be a reason to add the \( D \) branes to the ten
dimensional action, as GKP do, and not the anti-branes. However,
if both the branes and anti-branes are treated in the same manner
the classical potential becomes
\begin{equation}
 \label{potential2}
V=\int d^{6}y\sqrt{\tilde{g}^{(6)}}\frac{e^{4\omega (y) -
12u(x)}}{24\tau_{I}}
\widetilde{|iG_{3}-*_{6}G_{3}|^{2}}+2e^{-12u(x)}
\sum_{\overline{D}}T_{3}e^{4\omega (y^{\overline{D}})}.
\end{equation}
Note that $T_3$ is the brane tension, not to be confused with the
gaugino condensation field discussed in the previous section or
with the $T$ modulus. The contribution of the anti-D branes is
local and therefore their contribution is determined by the warp
factor at their positions $y^{\overline{D}}$.

If we follow KKLT and integrate out (classically) the complex
structure moduli and the dilaton we are left with an effective
four dimensional theory with a potential for the volume modulus
\begin{equation}
\label{dbarpot} V=\frac{2}{T_{R}^{3}}\sum
_{\overline{D}}T_{3}e^{4\omega (y^{\overline{D}})}.
\end{equation}

However, this is not a four dimensional SUGRA theory any more.
From the four dimensional stand point SUSY is explicitly broken by
the anti-D branes, as is evident by the term (\ref{dbarpot}) in
the potential. Moreover, it is a runaway potential which pushes
the theory towards the decompactification limit \(
T_{R}\rightarrow \infty \). In this limit ten dimensional SUSY
will be restored. This behavior is reminiscent of what happens
with the Scherk-Schwarz mechanism where a runaway potential is
generated for a modulus, though in that case the sign is opposite
to that in the above.

Since in the resulting four dimensional theory SUSY is explicitly
broken, it is no longer possible to derive the moduli potential
from a superpotential.  Hence it is unclear how the addition of a
non-perturbative contribution (coming say from gaugino
condensation) to the GVW superpotential evaluated at the minimum
of the complex structure moduli and the dilaton potential, can be
justified. A possible consistent derivation would be
possible if the term (\ref{dbarpot}) can be interpreted
as a D-term. However it is unclear how this can be done in this case. In
particular the D-term breaking discussed in
BKQ \cite{Burgess:2003ic} has a very different structure.

In BKQ SUSY is broken by turning on an electric flux \( E \) on \(
D7 \) branes which can be interpreted in \( \cal N \) =1 SUGRA
context as a U(1) D-term. This then gives (after integrating out
the complex structure moduli and the dilaton) a potential of the
standard D-term form \[ V=g^{2}_{YM}\frac{D^{2}}{2}=\frac{2\pi
}{T_{R}}\left( \frac{E}{T_{R}}+\sum q_{I}|Q_{I}|^{2}\right)
^{2},\] where the \( Q_{I} \) are any additional massless matter
fields which are charged under the gauge field on the \( D7 \)
branes with charges \( q_{I} \).

There are some uncertainties in this scenario. Generically,
charged massless matter exists and acquires vacuum expectation
values so as to set the D term to zero. This is exactly what was
realized in the context of the heterotic string (see for example
the discussion in section 18.7 of \cite{Polchinski:1998rr}).
However, in \cite{Burgess:2003ic} it was argued that there are
special situations in which such massless matter is absent, and
the D-term contribution is not cancelled. This seems to require a certain open
string  modulus to be fixed at a non-zero value but it is unclear how this
is done. In
addition, it should be noted that this scenario implies the
existence of an anomaly in the U(1) group since the D-term
interpretation of the \( E^2/T_{R}^{3} \) term depends on gauging
the PQ symmetry associated with $T_I$, the axionic partner of \(
T_{R}. \) This anomaly would need to be cancelled by some chiral
fermions having \( trQ\neq 0 \). The models of
\cite{Burgess:2003ic} do not have such fermions,
though it is possible that such models can be constructed.

\subsection{Racetrack models for Kahler moduli}

In view of the uncertainties associated with the proposals that we
have just described, we will examine another possibility, which in
some sense is a more conservative one, for finding a
positive or vanishing minimum as well as stabilizing the volume
modulus. We will consider field theoretic non-perturbative effects
in the superpotential incorporating multiple gaugino condensates
in the spirit of the old ``race-track" models, thus generalizing
the analysis of KKLT.

If we use only the ingredients of GKP without the anti-branes and
without turning on fluxes on the branes, the resulting effective
four dimensional theory is an \( \cal N \)=1 SUGRA. In this case
it is meaningful to add  non-perturbative contributions to the
superpotential.  Multiple gaugino condensates arise when the gauge
group is broken by turning on  discrete Wilson lines on the four
cycle of the CY manifold which is wrapped by the $D7$ branes. Note
that we have added another discrete choice to the discretuum. The
additional layer increases the number of vacuua in a way that
depends on the gauge group and the desired pattern of breaking.

It is likely that there will be some corrections to the Kahler
potential which we will ignore for the moment, since as long as
they are small their exact form is not particularly important to
us. The SUGRA potential obtained with all the ingredients
mentioned above is not necessarily a non-negative potential, and
it may have both positive and negative minima. There does not seem
to be a general argument which says that positive or vanishing
minima are somehow excluded. To rule out the positive minima on
the basis of a generalization of the classical no-go theorem would
require one to show that  there is a ten dimensional action which
incorporates the non-perturbative terms \emph{and} satisfies the
strong energy condition. As far as we know such an action does not
exist, and a priori there is no reason to dismiss the possibility
of a dS or Minkowski minima for the \( \cal N \) =1 SUGRA
potential in the case that we have discussed.

As in the heterotic case, there are two possibilities which are
feasible to analyze:

\begin{itemize}

\item The one light modulus case: Here we integrate out
(classically) the complex structure moduli and the dilaton both of
which generically will have string scale masses. Then by adding
the gaugino condensate terms to the constant superpotential coming
from the flux we are left with a potential for $T$. With one
condensate it is easy to see that the only minimum is an anti
deSitter (AdS) one, and that SUSY is preserved. On the other hand,
with more than one condensate one might have expected to find dS
or flat space minima. We will show that this is impossible
\footnote{A particular case of this general result was noticed in
\cite{Escoda:2003fa}.}.
 \item Two light moduli. With non- generic fluxes it
should be possible to keep the dilaton light while the complex
structure moduli will still have string scale masses.
Alternatively we could consider compactification on a CY with \(
h_{11}=2 \) and proceed as in the previous case after integrating
out the dilaton and the complex structure moduli at a high scale
and getting a superpotential for the Kahler moduli from gaugino
condensation. In this case we will have a \( \cal N \)=1 theory
with two light moduli and then, as we will show, it is possible to
have dS or Poincare minima.
\end{itemize}

\section{Poincare and de Sitter Minima of one and two moduli potentials}

We will be considering the minima of the SUGRA action
(\ref{vsugra}) where all but one or two moduli have been
integrated out at a high scale. The origin of the SUGRA action can
be either Type IIB string theory or the heterotic string theory.

In the type IIB string theory compactified on a CY manifold with
only one Kahler modulus the situation analyzed in the literature
(for instance KKLT) would fall into the category of the one
modulus case since the dilaton and the complex structure moduli
have been integrated out classically. The classical Kahler
potential of the SUGRA is
 \begin{equation}
\label{k3ln}
 K=-3\ln (T+\overline{T}),
 \end{equation}
and the  superpotential is of the form
 \begin{equation}
 \label{rctrack1}
  W=c+\sum d_{i}e^{-8\pi ^{2}T/C(G_{i})}.
 \end{equation}
The constant \( c \) in the superpotential is the value of the GVW
superpotential \cite{Gukov:1999ya} evaluated at the point that
minimizes the classical superpotential, and \( d_{i}=-\mu
^{3}C(G_{i}). \) In the type IIB case the sum originates from
multiple gaugino condensates that may occur if the original gauge
group living on the D7 branes is broken by discrete Wilson lines
on the four cycle which is wrapped by the branes.  The sum of
exponential terms in the superpotential would be over the simple
gauge group factors.

In the heterotic string theory case a similar effective four
dimensional SUGRA action for a single field can arise if (as is
generically the case) the dilaton and the complex structure moduli
would get string scale masses from (\ref{vsugra2}), and a
potential for the volume modulus arises from  the mechanisms
discussed in sect.~II.

The two light moduli case can arise in both string theories. One
possibility is that the overall volume modulus and the dilaton are
light. This can happen by tuning the parameters of the 10D action
by a choice in the discretuum. Other possibilities can arise as
well. For example, as an alternative to keeping the dilaton and
the overall volume modulus light, we may consider compactification
on a CY manifold with two Kahler moduli \( T_{1},\, T_{2} \) say.
Now we will have two four cycles (each assumed to have \( b_{1}=0
\) so that there are no open string moduli) and it is possible
have a stack of seven branes wrapping each cycle. We do not know
how to parameterize the corresponding metric in the CY case but we
might proceed in analogy with the torus (or orbifold) case where
there will be three Kahler moduli. After fixing the complex
structure moduli the metric may be written as,
\[
ds^{2} = e^{-2\sum ^{3}_{i=1}u_{i}(x)}
\tilde{g}_{\mu\nu}(x)dx^{\mu }dx^{\nu } +
e^{2u_{1}(x)}dz^{1}d\overline{z}^{1} + e^{2u_{2}(x)} dz^{2}
d\overline{z}^{2} + e^{2u_{3}(x)} dz^{3} d\overline{z}^{3}.\] We
need to identify the axionic partners of the Kahler moduli in
order to complete the chiral scalars in the corresponding
supermultiplets. They can be identified by writing the four form
gauge field
 as \( C_{4}=\sum_{i=1}^{3}a^{i}(x)\wedge J^{i} \) where \( J^{i}=dz^{i}\wedge
d\overline{z}^{i} \)  and the \( a^{i} \) are
two forms in four dimensions. The axions are then the
pseudo-scalars \( b^{i} \) defined by writing \(
da_{2}^{i}=e^{-2\sum _{j}u^{j}-2u^{i}}\tilde{*}_{4}db^{i} \) and
the chiral scalars in the Kahler moduli superfields take the form
\( T^{i}=\frac{b_{i}}{\sqrt{2}}+ie^{4u_{i}} \) with the Kahler
potential being \[ K=-\sum _{i=1}^{3}\ln
(T^{i}+\overline{T}^{i}).\]

In this case we can have three stacks of \( D7 \) branes each
wrapping a different four cycle and then with a certain choice of
normalization of the integral over the three cycles, we get for
the gauge coupling functions, \[ f^{1}=\sqrt{T^{2}T^{3}},\,
f^{2}=\sqrt{T^{3}T^{1}},\, f^{3}=\sqrt{T^{1}T^{2}}.\] The
corresponding VY superpotential would then be
\[
W=c+d_{i}e^{-8\pi ^{2}f^{i}/C(G_{i})}.\] Again, if one introduces
discrete Wilson lines on each four cycle then each exponential
term would be replaced by a sum of exponentials as before. The
point is that now we have a theory of three light moduli. As long
as the extrema of the potential are away from zero \footnote{In
any case the theory breaks down for values of $ T^{i} $ smaller
than unity, corresponding to compactification scales smaller than
the string scale.}, we can expand around any of them as before to
determine whether they are minima. However, the three moduli case
is technically quite complicated to analyze since the number of
terms in the potential is large. We wish to consider a simpler
case with only two light moduli, in other words we need the
analogous theory when the compactification manifold is a CY with
\( h_{11}=2 \). It is not clear to us how to compute the gauge
coupling functions in this case. However, all that we really need
is that the gauge coupling functions \( f^{1},\, f^{2} \) coming
from branes wrapping different four cycles have different
dependencies on the two moduli. By analogy with the case of the
torus  (with, say, \( T^{2}=T^{3} \)) this would appear to be the
case. If so, this case can be analyzed in exactly the same way as
the one with the dilaton and the overall volume.

In \cite{Brustein:2001ci} we have considered the constraints on a
four dimensional SUGRA with stable moduli if hierarchically small
SUSY breaking is desired, with an acceptably small CC. More
precisely, defining the ratio of the gravitino mass to the Planck
mass $\varepsilon=m_{3/2}/M_{Pl}$ the SUSY breaking is \(
O(\varepsilon ) \) and the CC is \( \ll O(\varepsilon ^{2}) \). We
showed that for general Kahler potentials, the mass of the modulus
in the SUSY breaking direction had to be \( O(\varepsilon ) \),
however the  masses of the other moduli could be of the string
scale. Assuming a canonical form of the Kahler potential we found
concrete examples of a one modulus potential with a stable
minimum. Our result does not preclude the existence of more than
one light modulus - it just requires at least one.

In earlier work (see \cite{Brustein:2000mq} and references
therein), it was found that single field steep superpotentials do
not allow extrema with broken SUSY and a non-negative CC that are
true minima in the resulting SUGRA potentials. Steep
superpotentials are defined by the condition that their
derivatives are large,
\begin{equation}
 \frac{\left|(T+\overline{T})\partial_T^{(n+1)} W\right|}
 {\left| \partial_T^n W \right|}\gg 1\ \ \
 n=0,1,2,3.
  \label{steepness}
  \end{equation}
The reason that $n \le 3$ appears will be explained shortly.

The steepness property holds for all the gaugino-condensation
superpotentials that were previously discussed, and it is generic
to all models of moduli stabilization near the boundaries of
moduli space. The typical example of a superpotential satisfying
(\ref{steepness}) is a sum of exponentials $W(T)=\sum\limits_i d_i
e^{-\beta_i T}$, as in eq.(\ref{rctrack1}),  in the region
$|T\beta_i| \gg 1$. In general, the precise definition of the
region in which the potential is steep will depend on its detailed
properties. Our result did not depend on the particular form of
the Kahler potential, provided that it was regular at the
extremum. Thus, to obtain a true minimum with broken SUSY and a
vanishing or positive CC requires that the superpotential and its
first three derivatives can be tuned so that the conditions in
eq.(\ref{steepness}) are avoided in a certain region of field
space.

We have further defined a criterion for what constitutes an
acceptably small CC: that the value of the CC be much smaller than
$\varepsilon^2 M_{Pl}^4$. The reason for choosing such a
criterion, and not requiring that the cosmological constant
vanishes or is of the order of the critical energy density as
suggested by recent observations is that we expect corrections to
the CC coming from low energy field theoretic effects.
Generically, loop contributions to the CC can be as large as
$\varepsilon^2 M_{Pl}^4$, and in addition we expect
contributions of order $\varepsilon^4 M_{Pl}^4$ from electroweak
scale physics. There are models in which the leading order
loop corrections can be cancelled (see for example \cite{Ferrara:1994kg})
but there would be corrections that are at least as large as
$\varepsilon^4 M_{Pl}^4$.
Therefore at energies just below the string scale,
a model with a negative CC that is as large as, say,
$\varepsilon^4 M_{Pl}^4$, is at the level of accuracy that we (and
others) are working at, as good a model as one with a positive CC
whose magnitude is $(10^{-3}eV)^4$.

Here we extend and complete our previous results. We will prove
the following results:
\begin{itemize}

\item If the Kahler potential for a single modulus $\Phi$ is of
the form that comes from classical string theory $K=- A \ln
(\Phi+\overline{\Phi})$, for $1\le A\le 3$, for example, \(
K=-3\ln (T+\overline{T}) \) or \( K=-\ln (S+\overline{S}) \), then
there does not exist a minimum with a positive or zero CC for an
F-term potential for any superpotential. A minimum with a negative
CC and broken SUSY in a region in which string perturbation theory
is under control can be found for such Kahler potentials, and with
enough tuning of the parameters of the superpotential it can be
made acceptably small.

\item In the case of two moduli (say, \( S \) and \( T \)) with a
classical string theory Kahler potential, for example \( K=-\ln
(S+\overline{S})-3\ln (T+\overline{T}) \),  it is possible to find
a superpotential that yields a stable minimum with a positive or
vanishing CC. It is also possible to find minima with a negative
CC and broken SUSY which, with enough tuning, can be made
acceptably small.

\end{itemize}

The first result means that if all but one modulus is integrated
out at a high scale, leaving us with an effective four-dimensional
SUGRA as in the KKLT case, then such a potential (without
introducing D-terms) cannot have a minimum with a zero or positive
CC. As we have argued above the situation with D-terms is unclear
so in our view the existence of such a dS or even Poincare minimum
is still not established. Our result also explains why the
race-track models with one modulus that have been discussed in the
literature over the past fifteen years have failed to produce a
model for stabilizing moduli with zero (or positive) CC.

The second result establishes the possible existence of a minimum
in the region of moduli space where the Kahler potential is
approximately of its classical form if all but two moduli are
integrated out at a high scale. In this case the effective
low-energy theory is a four-dimensional SUGRA with two light
moduli. We expect that the tuning that is available
in the discretuum will be sufficient to yield such a potential.

To establish our results we will begin with some general
considerations. We investigate a SUGRA potential given by
(\ref{vsugra}) with Kahler potential \( K=-A \ln
(S+\overline{S})-B\ln (T+\overline{T}) \). Here $S$ and $T$ are
generic names for two moduli, for instance they could be the
dilaton and a Kahler modulus, or two Kahler moduli. Let us assume
that there is a extremum of \( V_{SUGRA} \) at \( S=S_{0} \) , \(
T=T_{0} \). Then near the extremum we may expand the
superpotential in a power series of the form
\begin{equation}
\label{wexpansion}
W(S,T)=\sum_{ij}a_{ij}(S-S_{0})^{i}(T-T_{0})^{j}.
\end{equation}

We may simplify the superpotential. If we make a translation
$S\rightarrow S+i {\rm Im} S_{0}$,  $T\rightarrow T+i {\rm Im}
T_{0}$, the Kahler potential is unchanged, so without loss of
generality we can take \( S_{0},\, T_{0} \) to be real. Since
$S,T$ are non-negative fields and in fact ${S_0}_R$ and ${T_0}_R$
have the meaning of a coupling or a geometric object such as a
radius of a cycle of the compact manifold, both of them are
positive. We may now rescale $S$ and $T$ by $S_0$ and $T_0$,
respectively to get,
\begin{equation}
\label{wexpansion1} W(\widetilde{S},\widetilde{T})=\sum_{ij}a_{ij}
S_0^i T_0^j (\widetilde{S}-1)^{i}(\widetilde{T}-1)^{j}.
\end{equation}
The Kahler potential is changed under this rescaling,
\begin{equation}
\label{rescaledkahler} K\rightarrow K(\tilde S,\tilde T)-
A \ln (2{S_0}_R) - B\ln(2{T_0}_R),
\end{equation}
which means that the potential is rescaled $V(S,T) \to
\frac{1}{(2{S_0}_R)^A (2{T_0}_R)^B}
V(\widetilde{S},\widetilde{T})$. If the extremum is a minimum,
this rescaling does not change its nature:A dS minimum remains a
dS minimum and a Poincare minimum remains a Poincare minimum.
Further simplification occurs because the superpotential appears
as an absolute value squared in the potential, and therefore if we
multiply it by a constant phase, the potential is unchanged. We
may use this freedom to set $a_{00}$ to be real. To summarize: we
may take the minimum to be at $S_0=1$, $T_0=1$, and the first
(constant) term in the superpotential to be real without loss of
generality.

To establish the existence of a Poincare or dS minimum we need to
establish the following.
\[
V|_{0}\geq 0,\, \partial _{S}V|_{0}=\partial _{T}V|_{0}=0,\, {\rm
Eigenvalues}\left[\partial ^{2}_{ij}V|_{0}\right]\geq 0,\] where
\( |_{0} \) means that the expressions are to be evaluated at \(
S=S_{0},T=T_{0} \) and the last expression is the Hessian matrix
of second derivatives of \( V \). Now from the holomorphicity of
the superpotential, we see from the form (\ref{wexpansion}) that
in calculating these quantities we need to keep only terms up to
the third order in \( S \) and \( T \) in \( W \). This is because
in the expression for the potential only \( W \) and its the first
derivatives appear. So for analyzing the existence of a minimum we
may limit ourselves to a superpotential of the form
\[
W=\sum _{i=0,j=0}^{i+j=3}a_{ij}(S-1)^{i}(T-1)^{j},\] with $a_{00}$
real. We will see later that to prove that a minimum cannot exist
it is sometimes possible to restrict this to  a quadratic
superpotential.

\subsection{The One Modulus Case}

\subsubsection{deSitter and Poincare minima are not possible}

Here we would like to prove that if the Kahler potential for a
single modulus $\Phi$ is $K=- A \ln (\Phi+\overline{\Phi})$, for
$1\le A\le 3$, for example, then there does not exist a minimum
with a positive or zero CC for an F-term potential for any
superpotential. We will prove our result by showing that under the
conditions stated above there is at least one direction in which
the extremum is a maximum rather than a minimum. Thus, we will
show that an extremum can be a saddle point or a maximum but not a
true minimum.

Rather than computing the Hessian directly and showing that under
the conditions mentioned above it has at least one negative
eigenvalue, namely that it is not a positive definite matrix, we
will show that $\frac{\partial^2 V}{\partial \Phi_R \partial
\Phi_R}+\frac{\partial^2 V}{\partial \Phi_I \partial \Phi_I} < 0$.
If this quantity is negative then the Hessian matrix cannot be
positive definite. This is because the condition that it is
positive definite is that its determinant and all of its principal
minors be positive. If $\frac{\partial^2 V}{\partial \Phi_R
\partial \Phi_R}+\frac{\partial^2 V}{\partial \Phi_I \partial
\Phi_I} < 0$ then  $\frac{\partial^2 V}{\partial \Phi_R
\partial \Phi_R}<0$ and/or $\frac{\partial^2 V}{\partial \Phi_I \partial
\Phi_I} < 0$, so at least one of the principal minors is negative.
There are two advantages to choosing this quantity as a
diagnostic, one is that it is a linear in the potential rather
than quadratic as the determinant. The second advantage is due to
the fact that $\frac{\partial^2 V}{\partial \Phi_R
\partial \Phi_R}+\frac{\partial^2 V}{\partial \Phi_I \partial
\Phi_I}= 4 \frac{\partial^2 V}{\partial \Phi
\partial \overline{\Phi}}$. Because the superpotential is a holomorphic function,
$\frac{\partial^2 V}{\partial \Phi \partial \overline{\Phi}}$ at
the extremum depends only on the superpotential and its first and
second derivatives at the extremum. Therefore we may use a
quadratic superpotential to analyze this quantity. This greatly
simplifies the analysis.

We consider a Kahler potential
\begin{equation}
\label{kahler1a}
 K=- A \ln (\Phi+\overline{\Phi}),\hspace{.5in}\hbox{$1\le A\le 3$},
\end{equation}
and a quadratic superpotential
\begin{equation}
\label{super1}
 W= a0+ (a1_R+i a1_I)(\Phi_R+i \Phi_I-1)+(a2_R+i a2_I)(\Phi_R+i
 \Phi_I-1)^2,
\end{equation}
which depends on 5 real parameters $\{a0,a1_R,a1_I,a2_R,a2_I\}$.
The expression for the resulting potential has many terms and it
is not practical to present it here. We use Mathematica to compute
it symbolically and manipulate it.

We then impose the condition that $V(1)=\epsilon\ge 0$, and that
$\Phi=1$ is an extremum $\partial_{\Phi_R}V|_1=0$,
$\partial_{\Phi_I}V|_1=0$. This results in 3 equations for the 5
parameters, leaving 2 of them free. We then compute
$\left[\frac{\partial^2 V}{\partial \Phi_R
\partial \Phi_R}+\frac{\partial^2 V}{\partial \Phi_I \partial
\Phi_I}\right]|_{\Phi=1}$ in terms of the remaining two parameters
and check whether there is a region of parameter space for which
it is positive. If we choose the two free parameters to be $a1_R$,
$a1_I$ we find that
\begin{eqnarray}
\left[\frac{\partial^2 V}{\partial \Phi \partial
\overline{\Phi}}\right]_{\hbox{$|_{\Phi=1}$}}&=& -
\frac{2^{3-A}}{A(3-A)} \Biggl[ (3-A) a1_I^2 + (3+A) a1_R^2
\nonumber \\ &+& \sqrt{A} a1_R \sqrt{12 a1_R^2 - (3-A) \left(2^A A
\epsilon - 4 a1_I^2\right) }\Biggr] \hspace{.3in} A \ne 3
\nonumber \\
\left[\frac{\partial^2 V}{\partial \Phi \partial
\overline{\Phi}}\right]_{\hbox{$|_{\Phi=1}$}}&=& -2 \epsilon
\hspace{3in} A = 3.
 \label{egnvl1}
\end{eqnarray}
We then check whether this expression can be positive for any
value of $a1_R$ and $a1_I$ and we find that it is always negative.
This is of course obvious for the $A=3$ case.

If we look specifically for a Poincare vacuum, for which
$\epsilon=0$, the analysis simplifies, and the results remain the
same, the Hessian has  always at least one negative eigenvalue.
For the case $A=3$ it is less obvious, but it is nevertheless
correct since in this case we have from the above,
 $\left[\frac{\partial^2 V}{\partial \Phi_R
\partial \Phi_R}\right]_{\hbox{$|_{\Phi=1}$}}=
- \left[\frac{\partial^2 V}{\partial \Phi_I \partial
\Phi_I}\right]_{\hbox{$|_{\Phi=1}$}}$.

\subsubsection{AdS minima with SUSY breaking}

We would like to show that it is possible to find AdS minima with
an acceptably small CC. As can be seen from eq.(\ref{egnvl1}), if
the value of potential at the minimum is negative $\epsilon<0$, it
is no longer possible to deduce that one of the eigenvalues is
negative. In fact, it is rather easy to find examples when both
eignevalues are positive and therefore the candidate extremum is
indeed a minimum.

For example, in the case of a quadratic superpotential with real
coefficients $W(\widetilde T)=a_{0}+a_{1} (\widetilde
T-1)+a_{2}(\widetilde T-1)^2$, and a Kahler potential $K=-3
\ln(T+\overline T)$, the conditions for a true minimum are that
$a_{1}=a_{2}$, $a_{0}=\frac{a_2}{3}-\frac{2\epsilon}{a_2}$ and $0<
a_{1} < \sqrt{-6\epsilon}$. SUSY is generically broken at the
minimum since $F=-\frac{3}{2}a_0+a_1= a_2/2+3\epsilon/a_2$
generically does not vanish. For the case $K=- \ln(T+\overline
T)$, the conditions are $a_{0}=4 a_{2}$, $a_{1}=2a_{2}+{\sqrt{12
a_{2}^2 +\epsilon/2}}$, and at the minimum SUSY is generically
broken $F=-\frac{a_0}{2}+a_1=\sqrt{12 a_2^2+\epsilon/2}$.

To discuss the issue of scaling let us explicitly present the
relationship between exponential and polynomial potentials.
Comparing $W=c+\sum d_{i}e^{-8\pi ^{2}T/C(G_{i})}$ to
$W(\widetilde{T})=\sum a_{i}(\widetilde T-1)^{i}$, we find that
$a_0=c +\sum d_{i}e^{-8\pi ^{2}T_0/C(G_{i})}$, $a_1=\sum
-d_{i}\beta_i T_0 e^{-8\pi ^{2}T_0/C(G_{i})}$, $a_2=\sum
d_{i}\frac{(\beta_i T_0)^2}{2} e^{-8\pi ^{2}T_0/C(G_{i})}$, and so
on.  One needs to tune the coefficients such that the
superpotential and its first three derivatives are of the same
order in a region in the vicinity of the minimum. To make the CC
acceptably small it needs to be less than $O(\varepsilon^2)$,
recall that $\varepsilon= m_{3/2} /M_{Pl}$. Therefore the
parameters in the superpotential need to be tunable to an accuracy
which is roughly $\varepsilon$ (recall that the true potential is
rescaled by a factor $(2 T_{0R})^3$).

The constant $c$ can be tuned by choosing parameters in the
discretuum, while the amount by which the other coefficients are
tunable is determined by the value of $T_0$ and $C(G_i)$. In the
type IIB case there seem to be enough possibilities to tune the
coefficients to the desired accuracy, while in the heterotic case
compactified on a Kahler manifold the amount of tuning seems to be
insufficient as long as the constraint that the rank of the group
be less than 22 exists.

\subsection{The Two Moduli Case: Examples with minima}

The two moduli case can be analyzed using the same methods as the
single modulus case. Because of the complexity of the analysis we
cannot give the results for the general case, rather we give some
specific examples where it is possible to find a true minimum with
a positive or vanishing CC and broken SUSY.

The examples that we were able to find : $ K=- \ln
(S+\overline{S})-3 \ln (T+\overline{T})$, and a general cubic
superpotential with 10 real coefficients,
\begin{eqnarray}
\label{super2}
 W &=& a0+ a1(S_R+i S_I-1) + a2_R(S_R+i S_I-1)^2 +
 a3 (S_R+i S_I-1)^3 \nonumber \\
 &+& b1 (T_R+i T_I-1)+
 b2(T_R+i T_I-1)^2+ b3(T_R+i T_I-1)^3 \nonumber \\
 &+& ab1(S_R+i S_I-1)(T_R+i T_I-1)+
 ab2(S_R+i S_I-1)(T_R+i T_I-1)^2 \nonumber \\
 &+& ba2(S_R+i S_I-1)^2(T_R+i T_I-1).
\end{eqnarray}
The conditions that $T=1$, $S=1$ is an extremum and that the value
of CC is $\epsilon$ constitute three equations leaving seven free
parameters. We were able to find true minima for a range of these
parameters such that the CC is positive or vanishing. For example,
$a1=1, b1=1, ab1=1, a3= 0.1, b3= -.1, ab2 = .1,
      ba2 = .1, \epsilon = .6$, or
$a1 = .8, b1 = .8, ab1 = .5, a3 =0.2, b3 = -.15, ab2 = .2, ba2 =
-.15, \epsilon =.3$. The range of parameters is finite. It is also
possible to find examples of true minima with vanishing or
positive CC for simpler superpotentials, for example a general
quadratic superpotential with real coefficients.

For other forms of the Kahler potential we were unable to find
solutions with a positive or vanishing CC, however, the analysis
seems more complicated and our inability to find such solutions
does not necessarily mean that they do not exist. We do believe
that the existence of such solutions is quite sensitive to the
form of the Kahler potential. From the previous discussion it is
also clear that in addition to the solutions that we have found
for the particular form of the Kahler potential it is also
possible to find minima with a negative CC and broken SUSY, and
that with enough tuning their CC can be made small enough so that
they are acceptable according to the criteria that we have defined
previously. The amount of fine tuning required can be estimated by
comparing the polynomial form of the superpotential to its
original form as a sum of a constant and exponential terms as was
done in the single field case.

\section{Cosmological issues and conclusions }

\subsection{The Discretuum and the Anthropic Principle}

In the last few years it has become fashionable to apply the
Anthropic Principle (AP) to the discretuum. Imagine applying it to
the larger set of vacua that we have discussed. In order to
discuss the applicability of AP in string theory we first need a
precise statement of it. A scientifically acceptable statement
would be the following weak form of the AP:

\begin{itemize}
\item AP: Given a theory which predicts a range of values for some
fundamental parameters, observers will measure values for these
parameters that are typical of those universes which are
consistent with the existence of the observers.

\end{itemize}
To formulate this more precisely would require one to know exactly
what parameters of the standard model are relevant for our
existence. Weinberg \cite{Weinberg:1988cp, Martel:1997vi} has
argued that if the CC is not within a factor of a few of the
currently observed value then galaxies would not have formed (and
hence we would not have come into being). This argument may
constitute an explanation of the coincidence problem (why the CC
is of the same order of magnitude as the matter density), however,
it is not an explanation of its actual value. It is possible to
imagine universes where both the CC and the matter density are
much higher but galaxies are formed \footnote{A detailed analysis
along these lines is in \cite{Aguirre:2001zx}.}.

As observed by many authors the Anthropic Principle makes sense
only in the context of a theory which allows a wide range of
values for the parameters in question. The example of Newtonian
dynamics which allows for the existence of planets at various
distances from the sun is often cited. There this is just a matter
of a set of initial conditions, chosen presumably at random, with
one of them putting a planet just at such a distance that its
surface temperature is between the freezing point and boiling
point of water, so that life as we know it can form. In this
analogy the point is that there is no fundamental explanation of
why the planet Earth is at a certain distance from the sun. If it
was at a different distance (outside some small range) then we
could not inhabit it. So in the same way we should not ask why it
is that we live in a universe with a particular value of the CC
since if it were different (again, outside some range) then we
would not be here to ask the question.

Notice that the argument assumes that all possible distances from
the sun are allowed and in principle can be achieved. If for
instance in spite of the theory, observation showed that there was
only one planet in the universe and it is at such a distance that
liquid water existed on it's surface, then the anthropic
explanation would not be tenable. Thus, the reality of planets at
other distances is a necessary condition for this explanation to
make sense. Similarly, an anthropic explanation of the
cosmological constant (or any other parameter) requires the
reality of other solutions to string/M theory that have different
values of this parameter. We can see that there are other planets
at a variety of different distances, but so far we have not
detected any other universe, and it is not clear that we ever
would \cite{Carroll}. The detection of other universes might even
be impossible in principle.

Thus the anthropic explanation actually entails a prediction -
that other universes exist, and that there is a correlation
between the values of their CC's and the existence of galaxies
capable of supporting intelligent life in them. The latter does
not make any sense unless the former is true. On the other hand,
for most of the history of fundamental physics, theories and
models that do not satisfy observational criteria have been
discarded as unphysical. Of course, in the past physicists have
not attempted anything as ambitious as the construction of a
theory of all fundamental phenomena. But an analogy from General
Relativity may illustrate the point. As is well known, without
additional criteria this theory can lead to bizarre solutions,
including naked singularities universes with closed time-like
geodesics, etc. \footnote{Recent work seems to indicate that Goedel
universes are valid solutions of string theory too.}. Even
solutions which are much more acceptable such as say a Bianchi
cosmology are usually rejected because they are not in accord with
the observed homogeneity and isotropy of the universe.

Application of the anthropic principle in the discretuum would
make sense only if one were to treat all possible solutions of
perturbative string theory as having a real existence. Since it is
highly unlikely that any of them other than our own (assuming it
is a member of the discretuum) is ever going to be observed it
does not appear to be a meaningful principle to use.
 In this paper we have argued that there are points in the discretuum
where all the moduli are stabilized. Perhaps it is possible to
find such a point with the observed CC and small supersymmetry
breaking. If so, rather than appeal to the anthropic principle, we
would argue that what has been done is good old fashioned
fine-tuning!

\subsection{The overshoot problem}

A generic problem with the moduli stabilization and SUSY breaking
scenarios discussed above is that cosmologically they are all
subject to the overshoot problem first pointed out in
\cite{Brustein:1993nk}. To see this let us estimate the height of
the barrier separating the SUSY breaking minimum from the runaway
decompactifying region of the potential. In both the KKLT and the
BKQ proposals this is clearly set by the size of the extra term
that is added. In either case at the minimum this number sets the
scale of SUSY breaking. To have low energy SUSY the D term in the
BKQ proposal should satisfy \( |D|<O(10^{-14}) \) in Planck (or
string) units. In the anti-D brane case the SUSY breaking is
explicit but if one needs this breaking to be at a low scale then
we will have a similar result. As one moves away to the right of
the minimum, this term gets reduced (since it is proportional to a
negative power of \( T_{R} \)) and hence the barrier height is \(
\leq |D|^{2}\sim 10^{-28} \). In the case of F-term breaking (with
multiple gaugino condensates) discussed here the argument is
slightly more involved. The point is that at the SUSY breaking
minimum, in order to get a nearly zero CC, the value of \(
|\sqrt{3}W| \) should be equal to \( |F| \) i.e the value of the
SUSY breaking order parameter. At this point then the value of the
exponential terms is of the same order as the constant in the
superpotential. Beyond this minimum the exponential terms are
smaller (in absolute value) and an extremum at a positive value of
V may arise but the barrier is expected to be of the same order as
\( |F|^{2} \) at the SUSY breaking scale.

In addition to the SUSY breaking minima that we have worked so
hard to establish, in generic situations there are also nearby
SUSY preserving minima with a large negative CC. The reason is
that for superpotentials that can be approximated by polynomials
(and this can always be done near the SUSY breaking minimum) there
are also solutions to the equations $F=0$. Whenever the $F$-term
vanishes the potential is negative (or in some special situations
vanishes), and in general one of these solutions will be the
global minimum in the region where the approximation holds.

Now the problem is that generically one would expect the initial
conditions on the \( T \) modulus to be set by the string/Planck
era of the universe when one expects string scale energy densities
and temperatures. Clearly if the modulus starts with energy
density \( \leq O(1) \) in string units then it is not going to
remain in this extremely shallow minimum and will roll right over
into the decompactifying region or into the SUSY preserving AdS
minima. This classical rolling rather than quantum tunnelling
through the barrier is the real problem with any cosmology based
on such outer region compactifications. Of course it is possible
(though unlikely) that with enough tuning of the  parameters such
that the height and width of the barrier are much larger, some of
these problems may be avoided, but in the absence of a concrete
example one has to regard this issue as a serious problem.

\subsection{Towards a resolution}

The main focus of this paper has been the possibility of obtaining
models with all moduli stabilized. We have established the
following for \( \cal N \)=1 SUGRA potentials with the classical
string theory form for the Kahler potential, and a positive or
zero CC:

\begin{itemize}

\item If all but one modulus is stabilized at a high scale then it is not
possible to have the remaining light modulus stabilized by F terms.

\item If there are two light moduli then there are examples where
stabilization can be achieved in regions where string perturbation
theory is under control.

\end{itemize}

Our results depend on the form of Kahler potential, as we have
emphasized on several occasions along the way. If the corrections
to the classical form of the Kahler potential are small, as
expected in regions of moduli space in which string perturbation
theory is a good approximation, then our results should be valid.
If for the scales at which the CC is measured (or even at the
standard model scale) it will turn out that there are significant
corrections to the Kahler potential and the classical form is
drastically modified, then we expect our results to be
significantly modified. For example, if the Kahler potential is
modified in such a way that it can be approximated by the
canonical form, then one can find a good minimum even in the one
modulus case \cite{Brustein:2001ci}.  Clearly for two moduli there
is a wider range of possibilities.

The minima that we have discussed, as well as all others discussed in
the literature, fall into the category of outer region solutions
in the terminology of \cite{Brustein:2000mq}. They are separated
from the runaway regions of the moduli potentials by tiny barriers
and are thus subject to the cosmological overshoot problem
discussed in the previous subsection. This means that even if one
finds a model of this sort which contains the (supersymmetric
extension of the) standard model, such a theory - though it would
serve as an existence proof that an ultra-violet completion of the
standard model coupled to gravity exists - would not give a viable
cosmology.

If one includes the requirement of a viable cosmology it appears
unlikely that one could get a satisfactory theory in the outer
region of moduli space. Thus, as has been the recurrent theme of
our previous work,  we need to focus on building models in the
central region of moduli space - i.e the region which is not
related by any dualities to weak coupling large volume
compactifications. Obviously, it is technically hard to compute in
this region at this stage of development of string theory and
perhaps one has to await the successful formulation of some
non-perturbative description of string theory to be able to
calculate meaningful quantities in this region. However, as
explained in \cite{Brustein:2000mq} and in \cite{Brustein:2001ci}
combining information from the different theories in the outer
region and using information from a bottom up approach one may
gain some insight into the physics of this region.

\section{acknowledgements}
This research was supported by grant No.~1999071 from the United
States-Israel Binational Science Foundation (BSF), Jerusalem,
Israel.  SdA was also supported in part by the United States
Department of Energy under grant DE-FG02-91-ER-40672. R.~B. thanks
the KITP, UC at Santa Barbara, where this work has been completed.
We would like to thank  the participants of the ``superstring
cosmology" program at KITP for many discussions on issues relevant
to this paper, and in particular  T. Banks, C. Burgess, M. Dine,
S. Kachru, R. Kallosh, A. Linde, J. Polchinski, F. Quevedo, and M.
Schultz. We would like to thank E. Novak for collaboration during
the early stages of the project.


\begin{thebibliography}{10}

\bibitem{Derendinger:1985kk}
J.~P.~Derendinger, L.~E.~Ibanez and H.~P.~Nilles,
Phys.\ Lett.\ B {\bf 155}, 65 (1985).

\bibitem{Dine:1985rz}
M.~Dine, R.~Rohm, N.~Seiberg and E.~Witten,
\newblock Phys. Lett. {\bf B156}, 55 (1985).

\bibitem{Derendinger:1985cv}
J.~P.~Derendinger, L.~E.~Ibanez and H.~P.~Nilles,
Nucl.\ Phys.\ B {\bf 267}, 365 (1986).


\bibitem{Nilles:2004zg}
H.~P.~Nilles, ``Gaugino condensation and SUSY breakdown,''
arXiv:hep-th/0402022.

\bibitem{Krasnikov:jj}
N.~V.~Krasnikov,
Phys.\ Lett.\ B {\bf 193}, 37 (1987).

\bibitem{deCarlos:1992da}
B.~de Carlos, J.~A.~Casas and C.~Munoz,
Nucl.\ Phys.\ B {\bf 399}, 623 (1993) [arXiv:hep-th/9204012].




\bibitem{Witten:1996bn}
E.~Witten,
Nucl.\ Phys.\ B {\bf 474}, 343 (1996)
[arXiv:hep-th/9604030].


\bibitem{Bousso:2000xa}
R.~Bousso and J.~Polchinski,
\newblock JHEP {\bf 06}, 006 (2000), [hep-th/0004134].


\bibitem{Kachru:2003aw}
S.~Kachru, R.~Kallosh, A.~Linde and S.~P.~Trivedi,
Phys.\ Rev.\ D {\bf 68}, 046005 (2003) [arXiv:hep-th/0301240].

\bibitem{Burgess:2003ic}
C.~P.~Burgess, R.~Kallosh and F.~Quevedo,
JHEP {\bf 0310}, 056 (2003) [arXiv:hep-th/0309187].

\bibitem{Dine:1999dx}
M.~Dine and Y.~Shirman,
Phys.\ Rev.\ D {\bf 63}, 046005 (2001) [arXiv:hep-th/9906246].


\bibitem{Brustein:1993nk}
R.~Brustein and P.~J. Steinhardt,
\newblock Phys. Lett. {\bf B302}, 196 (1993), [hep-th/9212049].



\bibitem{dinepc}
M.~Dine, private communication.


\bibitem{Gukov:2003cy}
S.~Gukov, S.~Kachru, X.~Liu and L.~McAllister, ``Heterotic moduli
stabilization with fractional Chern-Simons invariants,''
arXiv:hep-th/0310159.


\bibitem{Rohm:1986jv}
R.~Rohm and E.~Witten,
\newblock Ann. Phys. {\bf 170}, 454 (1986).


\bibitem{Veneziano:1982ah}
G.~Veneziano and S.~Yankielowicz,
Phys.\ Lett.\ B {\bf 113}, 231 (1982).


\bibitem{Gukov:1999ya}
S.~Gukov, C.~Vafa and E.~Witten,
\newblock Nucl. Phys. {\bf B584}, 69 (2000), [hep-th/9906070].


\bibitem{Giddings:2001yu}
S.~B.~Giddings, S.~Kachru and J.~Polchinski,
Phys.\ Rev.\ D {\bf 66}, 106006 (2002) [arXiv:hep-th/0105097].


\bibitem{Brustein:2001ci}
R.~Brustein and S.~P. de~Alwis,
\newblock Phys. Rev. Lett. {\bf 87}, 231601 (2001), [hep-th/0106174].

\bibitem{Banks:1996ss}
T.~Banks and M.~Dine,
\newblock Nucl. Phys. {\bf B479}, 173 (1996), [hep-th/9605136].

\bibitem{Font:1990nt}
A.~Font, L.~E. Ibanez, D.~Lust and F.~Quevedo,
\newblock Phys. Lett. {\bf B245}, 401 (1990).

\bibitem{Kaplunovsky:1995jw}
V.~Kaplunovsky and J.~Louis,
\newblock Nucl. Phys. {\bf B444}, 191 (1995), [hep-th/9502077].

\bibitem{Strominger:1986uh}
A.~Strominger,
\newblock Nucl. Phys. {\bf B274}, 253 (1986).

\bibitem{Cardoso:2003sp}
G.~L.~Cardoso, G.~Curio, G.~Dall'Agata and D.~Lust,
 ``Heterotic string theory on non-Kaehler manifolds with H-flux and gaugino
condensate,'' arXiv:hep-th/0310021.


\bibitem{Becker:2003sh}
K.~Becker, M.~Becker, P.~S.~Green, K.~Dasgupta and E.~Sharpe,
Nucl.\ Phys.\ B {\bf 678}, 19 (2004) [arXiv:hep-th/0310058].



\bibitem{deAlwis:2003sn}
S.~P.~de Alwis,
Phys.\ Rev.\ D {\bf 68}, 126001 (2003)
[arXiv:hep-th/0307084].

\bibitem{Polchinski:1998rr}
J.~Polchinski,
\newblock ``String Theory", Cambridge, UK: Univ. Pr. (1998).

\bibitem{Escoda:2003fa}
C.~Escoda, M.~Gomez-Reino and F.~Quevedo,
JHEP {\bf 0311}, 065 (2003)
[arXiv:hep-th/0307160].

\bibitem{Brustein:2000mq}
R.~Brustein and S.~P. de~Alwis,
\newblock Phys. Rev. {\bf D64}, 046004 (2001), [hep-th/0002087].

\bibitem{Ferrara:1994kg}
S.~Ferrara, C.~Kounnas and F.~Zwirner,
Nucl.\ Phys.\ B {\bf 429}, 589 (1994)
[Erratum-ibid.\ B {\bf 433}, 255 (1995)]
[arXiv:hep-th/9405188].

\bibitem{Weinberg:1988cp}
S.~Weinberg,
Rev.\ Mod.\ Phys.\  {\bf 61}, 1 (1989).

\bibitem{Martel:1997vi}
H.~Martel, P.~R.~Shapiro and S.~Weinberg,
Astrophys.\ J.\  {\bf 492}, 29 (1998)
[arXiv:astro-ph/9701099].

\bibitem{Aguirre:2001zx}
A.~Aguirre,
\newblock Phys. Rev. {\bf D64}, 083508 (2001), [astro-ph/0106143].


\bibitem{Carroll}
S.~M.~Carroll, ``Why is the universe accelerating?,''
arXiv:astro-ph/0310342.





\end{thebibliography}

\end{document}